\documentclass[aps,prl,reprint,groupedaddress]{revtex4-1}
\usepackage{epsfig}
\usepackage{graphicx}
\usepackage{color}

\begin{document}

\title{Modeling vitreous silica bilayers}

\author{Mark Wilson}
\email[]{mark.wilson@chem.ox.ac.uk}
\affiliation{Department of Chemistry, Physical and Theoretical Chemistry Laboratory,
University Of Oxford, South Parks Road,Oxford OX1 3QZ, U.K}

\author{Avishek Kumar}
\email[]{akumar67@asu.edu}
\affiliation{Department of Physics,
Arizona State University, Tempe, AZ 85287-1604}

\author{David Sherrington}
\email[]{d.sherrington@physics.ox.ac.uk}
\affiliation{Rudolf Peierls Centre for Theoretical Physics,
University of Oxford, 1 Keble Road, Oxford OX1 3NP, U.K}
\affiliation{}
\affiliation{Santa Fe Institute, 1399 Hyde Park Rd., Santa Fe, NM 86501}

\author{M.F. Thorpe}
\email[]{mft@asu.edu}
\affiliation{Department of Physics,
Arizona State University, Tempe, AZ 85287-1604}
\affiliation{}
\affiliation{Rudolf Peierls Centre for Theoretical Physics,
University of Oxford, 1 Keble Road, Oxford OX1 3NP, U.K}

\begin{abstract}

 Theoretical modeling is presented for a free-standing vitreous silica bilayer which has recently been synthesized and characterized experimentally in landmark work.
While such two-dimensional continuous random covalent networks should likely occur on energetic grounds, no synthetic pathway had been discovered previously. Here the bilayer
is modelled using a computer assembly procedure initiated from a single layer of a model of amorphous graphene, generated using a bond switching algorithm from an initially crystalline {\it{graphene}} structure. Each
bond is decorated with an oxygen atom and the carbon atoms are relabeled as silicon,
generating a two dimensional network of corner sharing triangles. Each triangle is transformed into a tetrahedron, by raising the silicon atom {\it above} each triangular base and adding an additional singly coordinated oxygen atom at the apex. The final step
in this construction is to mirror-reflect this layer to form a second layer and attach the two layers to form the bilayer.
We show that this vitreous silica bilayer has the additional macroscopic degrees of freedom to form easily a network of identical corner sharing tetrahedra {\it{if}} there is a symmetry plane through the center of the bilayer going through the layer of oxygen ions that join the upper and lower monolayers. This has the consequence that the upper rings lie exactly above the lower rings, which are tilted in general. The assumption of a network of perfect corner sharing tetrahedra leads to a range of possible densities that we characterize as a {\it flexibility window}; with some similarity to flexibility windows in three dimensional zeolites.
Finally, using a realistic potential, we have relaxed the bilayer to determine the density
and other structural characteristics
such as the Si-Si pair distribution functions and the Si-O-Si bond angle distribution, which
are compared with experimental results obtained by direct imaging.
\end{abstract}


\maketitle

\section{Introduction}
The continuous random network model of network glasses is widely accepted as a model for materials like vitreous silica and amorphous silicon~\cite{Thorpe Wright}.

Although it is
more than eighty years since Zachariasen proposed this
model of glasses~\cite{Zach}, and experimental evidence has been compelling
over the years, especially through diffraction experiments~\cite{Thorpe Wright}, it has never been quite conclusive since the probability
distribution of rings of various sizes has been
elusive to determine explicitly experimentally.
This situation has now changed dramatically with the discovery and imaging~\cite{lichtenstein2012,huang2012}
of two dimensional bilayers of vitreous silica. Here, not only the distribution of rings, but the actual detailed atomic ring structure has been imaged for the first time in real space, removing all speculation from this subject (at least for this class of materials).
These are the first examples of which we are aware of real space imaging of a random network and as such represent
tours de force. Previously only small {\it{defect}} patches have been imaged, as for example for graphene as reported by Geim~\cite{Geim}.

In this paper, we provide the first atomic level computer model for a vitreous silica bilayer and demonstrate some intriguing and unexpected features that are shown to agree with experiment. There is a symmetry plane through the center of the bilayer where all the oxygen atoms that connect the tetrahedra in the lower and upper planes of the bilayer lie. Each tetrahedron comprises an SiO$_{4}$ unit and the whole bilayer is a perfect corner-sharing continuous random network with the
same chemical formula SiO$_{2}$ as $3d$ bulk vitreous silica.
Each monolayer is amorphous with rings from 4 up to
about 9 in size, consistent of course with Euler's theorem
that the average ring size is 6.
Because of the amorphous nature of the monolayer and the need for
oxygen bridges connecting the upper and lower layers,
it is necessary for the two layers to have the same ring structure and be {\em topologically} identical to form a complete corner sharing tetrahedral network. The result that the two layers are also {\em geometrical} mirror images of each other is quite surprising at first sight in a system that is {\it a priori} without any symmetry, but comes about from understanding the nature of the constraints within the network as explained later. This is consistent with our detailed atomic modeling and also is in accord with the experimental results~\cite{lichtenstein2012,huang2012} which show that the upper and lower layers do lie one on top of the other as required by a symmetry plane. We note that this does {\it not} imply that there is a three fold axis between the two upper and lower tetrahedral units (tetrahedral pair) through the common central oxygen  atom - rather this Si-O-Si angle through the central oxygen has a distribution of values throughout the sample, as do all the other Si-O-Si angles in the bilayer.

\begin{figure}[ht]
\centering
\includegraphics[angle=0, scale=0.30]{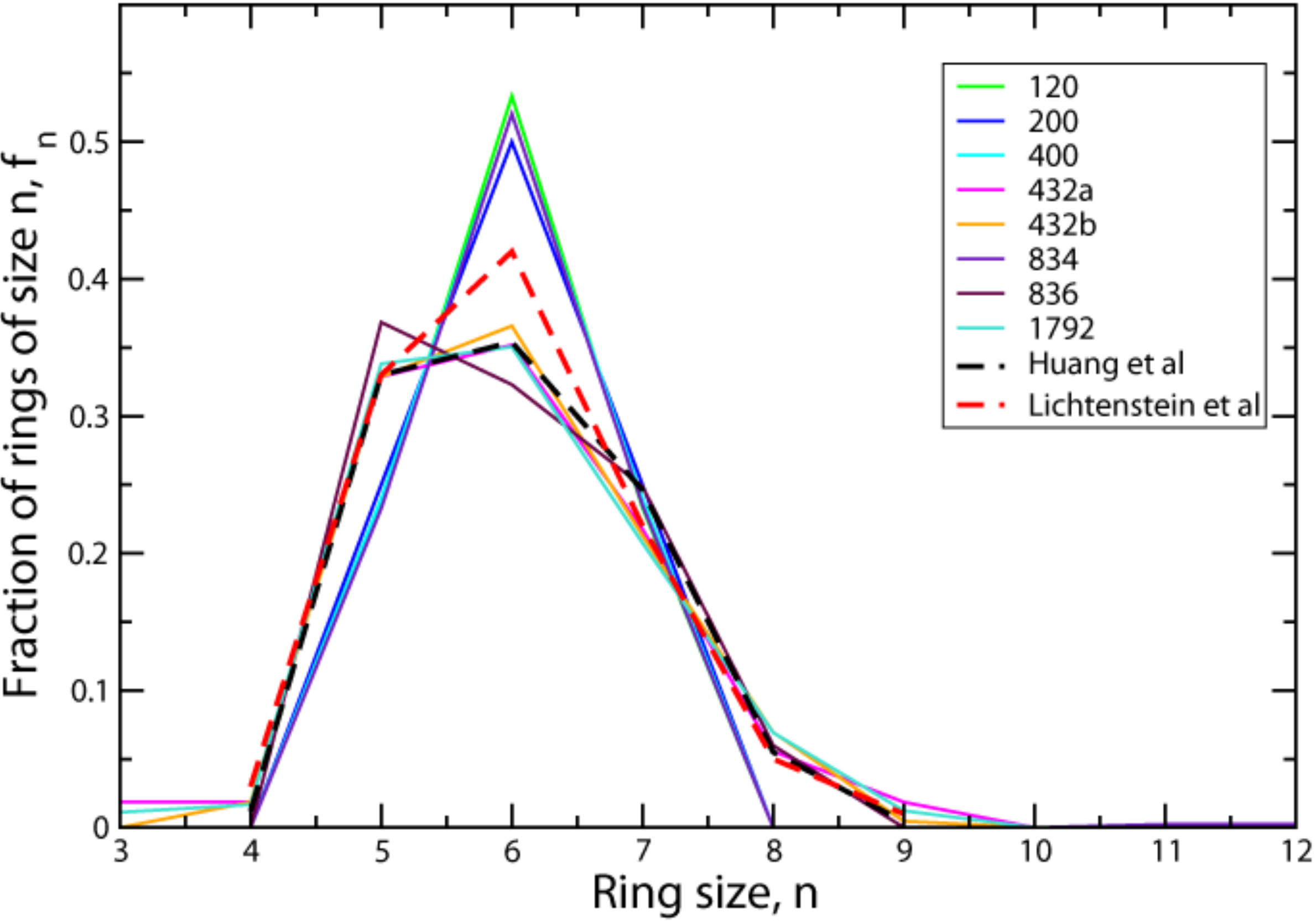}
\caption{Ring size distributions (fraction of rings containing $n$ silicon atoms)
obtained experimentally for vitreous SiO$_2$ bilayers
formed on Ru(0001) \cite{lichtenstein2012} - red dashed line, and graphene \cite{huang2012} - black dashed line.
Also shown are ring size distributions of computer-generated a-G seeds used in this study (with different generating protocols but labelled by sample size $N$).}
\label{fig:rings_expt}
\end{figure}

Thin vitreous SiO$_2$ films (interpreted as bilayers) have been grown on
Mo(112) \cite{weissenrieder2005}, Ru(0001) \cite{lichtenstein2012,heyde2012,loeffler2010}
and graphene \cite{huang2012}.
Figure~\ref{fig:rings_expt} shows the experimentally-obtained ring statistics
from references~\cite{lichtenstein2012} and~\cite{huang2012}.
A key observation is that the ring statistics obtained from the two experimental samples are not the same,
although this is not surprising in view of their
different preparation conditions, analogous to
various fictive temperatures used to characterize the preparation conditions for bulk silica~\cite{Thorpe5}.
\textcolor{black}{Different Monte Carlo annealing temperatures and/or protocols used computationally to create vitreous silica bilayers also lead to different ring statistics similar to those documented previously in amorphous cellular networks~\cite{Aste99} and in amorphous graphene~\cite{kumar2012}.}

What is unclear is
the extent to which
%
%
these differences in ring statistics
%
%
reflect the finite system sizes
under study or
%
%
the more complex
and more interesting
dependence of the structure on the precise
preparation conditions (including the nature of the substrate). The ring statistics are a fundamental quantity and
%
%
their
dependence on sample size imaged and on preparation conditions will be an important area for future study, especially experimentally.
The simplest non-trivial (second cumulant) measure of the
ring statistics
$\mu_2$ should be
 related to the static structure factor $S(0)$,
as
in bulk vitreous silica~\cite{Thorpe5}.
%

Recent simulation work, in which amorphous Graphene (a-G) configurations were generated using both {\it{bond-switching}}
Monte Carlo and molecular dynamics methods, highlights how networks constructed
primarily from 5-, 6- and 7-membered rings may adopt a range of structures \cite{kumar2012}.
A useful simple metric for distinguishing between the different samples is the second
moment of the ring size distribution,

\begin{equation}
\mu_2=(<n^2>-<n>^2),
\end{equation}
where $<n>$ is the mean ring size for an ideal two-dimensional network constructed from
purely three-coordinated sites ($<n>=6$ from  Euler's theorem). This metric conveniently captures the {\it{major}} changes in ring statistics from sample to sample in a single number. The values of $\mu_2$ for the experimental data presented
in Figure \ref{fig:rings_expt} are 0.904 and 0.886 (from References \cite{lichtenstein2012} and \cite{huang2012} respectively).

\section{Flexibility window}

\textcolor{black}{A key concept that will emerge is that there is a {\it {flexibility window}} involving $O(N)$ motions among the rigid corner-sharing tetrahedra.  This flexibility window designates a range of densities over which a framework of rigid tetrahedra, freely jointed at all corners with a given topology, can exist. The low density end of the window is defined by the maximum extension the framework can sustain without breaking apart, and the high density end of the window is determined by oxygen-oxygen overlap between adjacent tetrahedra. We will see that when additional terms are included in the potential, particularly the Coulomb terms, a particular density is selected from within the flexibility window. Similar ideas have been explored extensively in zeolites~\cite{Thorpe2} where the origin of the window is due to symmetry as in the vitreous silica bilayers studied here. However, in bulk zeolites the symmetry is associated with the rotations and translations of the unit crystallographic cell, whereas here the symmetry is due to a reflection symmetry, that is maintained between the two monolayers that comprise the bilayer.  We
will return to a full analysis of this latter point in a later section.
}

\section{Construction method}

\begin{figure}[ht]
\includegraphics[width=70mm]{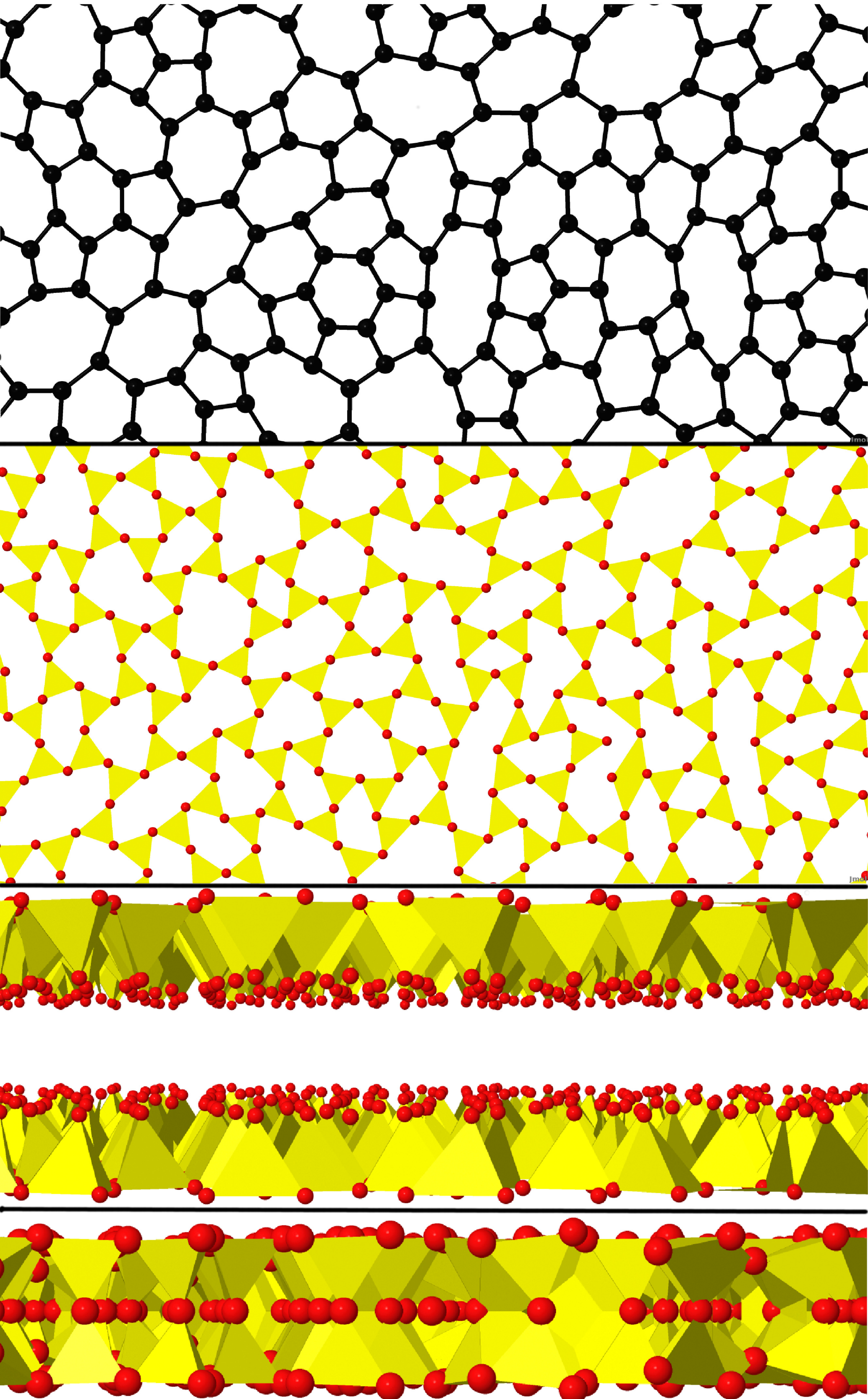}
\caption{The free standing vitreous silica bilayer is modeled by first creating an amorphous graphene layer from a crystalline graphene sheet through a bond switching algorithm (top panel), where the carbon atoms are shown as solid black dots. An oxygen ion is then placed at the center of each carbon-carbon bond and the carbon is {\it replaced} by a silicon ion forming a network of corner sharing triangles (second panel). The silicon atom is then lifted out of the plane and an oxygen ion (shown as a solid red dot) is placed above this silicon to form a tetrahedron. This silica monolayer is mirror inverted and placed directly above the first monolayer (third panel). Finally these two monolayers are brought together to form the silicon bilayer (fourth panel) in which the central oxygen ions are {\it combined} to make single bridging oxygen ions between the two monolayers.}
\label{fig:construction}
\end{figure}

The initial SiO$_2$ bilayer configurations are generated from {\it{ideal}} a-G coordinates (Figure \ref{fig:construction}), which were  themselves
generated using a ``bond-switching'' Monte Carlo algorithm (as described, for example,
in Reference \cite{kumar2012}; see also \cite{Aste99}). The a-G configurations generated in this manner are guaranteed to be constructed exclusively from three-coordinated carbon local environments. This method is superior to
others in the sense that it produces no coordination defects or dangling bonds, and is periodic with a super-cell
whose size can be chosen and varied.
\textcolor{black}
{Seed a-G configurations were constructed with a range of different ring statistics and hence $\mu_2$ values. Several different network sizes were employed. For convenience we shall refer to these systems below by their sizes, $N$  =120, 200, 400, 432 [two configurations, distinguished as (a) and (b)], 834, 836 and 1792 atoms. The two configurations containing 432 carbon atoms are generated with different ring statistics to give some extra perspective on the effect of the ring distribution on physical properties.}
Table~\ref{tab:mu2} lists the values of $\mu_2$ for these a-G configurations, while
Figure~\ref{fig:rings_expt} shows the detailed ring size distributions for these computer generated structures.\

\begin{table}[ht]
\begin{center}
\begin{tabular}{|c||c|}
\hline
$N$ & $\mu_2$ \\
\hline
120 & 0.467   \\
200 & 0.500    \\
400 & 0.480    \\
432a & 1.046    \\
432b & 0.935    \\
834 & 0.618    \\
836 & 0.856    \\
1792 & 1.014   \\
\hline
expt. {\protect{\cite{lichtenstein2012}}} & 0.904   \\
expt. {\protect{\cite{huang2012}}} & 0.886   \\
\hline
\end{tabular}
\end{center}
\caption{Variances, $\mu_2$, in the second
moment of the ring size distribution for the eight configurations
studied here, labelled by the number $N$ of atoms in the original graphene layers, or equivalently the number of Si atoms in a monolayer. The resulting bilayer therefore has $2N$ SiO$_2$ units. Also shown are the two experimentally-observed configurations.}
%
\label{tab:mu2}
\end{table}

The method for generating the bilayers is motivated by the observation that the two
layers sit on top of each other. As a result, each layer can be generated from the a-G configuration
and {\it{joined}} with oxide anion bridges. To generate the initial SiO$_2$ bilayer configurations, each carbon atom is {\it{transformed}} into
a silicon atom (which will eventually become the center of each SiO$_4$ tetrahedron). Oxygen atoms are
then
placed at the center
of each C-C bond to produce a single layer configuration of stoichiometry Si$_2$O$_3$
confined to (say) the $xy$ plane, which can be viewed as a two dimensional network of corner sharing {\it {equilateral}} triangles.
Each triangle has oxygen atoms at the vertices and a silicon ({\it transformed} from carbon) atom at the center. Additional oxygen  atoms are then placed perpendicular to the $xy$ plane
(which initially contained all of the atoms) at the center of each triangle and raised above to form a tetrahedron, with the silicon atom raised out of the plane to be at the center of the tetrahedron. This generates an Si$_2$O$_5$ network formed from
tetrahedra; each sharing three corners with a fourth corner unshared (for the moment) and with all unshared corners pointing {\it{up}}. The second layer of the bilayer is created by producing a mirror image
of the first layer (such that the tetrahedra are now pointing in the {\it down} direction) and offsetting the layer along the $z$-direction,
to lie above the first bilayer; see Figure 3(c).
Finally,
the
median oxygen atoms
%
%
 are {\it coalesced} between the two layers giving the required SiO$_2$ bilayer stoichiometry
[Fig~3(d)].
The system super-cell lengths
are then re-scaled so as to generate the required Si-O bond lengths. As a result, the systems
considered contain $2N$ [=240, 400, 800, 864(a), 864(b), 1668, 1672 and 3584] SiO$_2$ molecules.





\textcolor{black}{These structures are relaxed using standard Molecular Dynamics (MD) procedures with appropriate model inter-ion potentials.}
Two forms of potential model are considered. The first, which we will refer to as the {\it{harmonic potential}}, is designed to produce a corner sharing network of identical regular tetrahedra,
\textcolor{black}{with freedom of the individual tetrahedra to move and tilt  with respect to each other}
\textcolor{black}{while maintaining the topology.}
 This allows for hinging-freedom of the joined tetrahedral corners, and does not {\it{impose}} reflection symmetry.
A convenient way to accomplish this is to use  harmonic springs
to join the four nearest-neighbor Si-O and six nearest-neighbor O-O atoms in individual tetrahedron.
Computationally
%
%
%
the ratios of the O-O and Si-O equilibrium bond lengths of the potentials are chosen so as to produce ideal tetrahedra in isolation; thus the equilibrium separation for the neighboring oxygens along the edge of the tetrahedron in the O-O potential is taken to be
%
%
%
%
$\sqrt(8/3)\simeq 1.633$ times
that for
the Si-O nearest neighbor
separation.
For computational convenience the
spring force constants are taken to be equal
for both the Si-O and
O-O pairs within each tetrahedron. The detail of this interaction is only significant in the sense of allowing for a relatively
rapid energy minimization.
%
%

These
%
%
simple harmonic potentials do not however, preclude different tetrahedra from overlapping,
\textcolor{black}{as would be the case in reality, for example to prevent oxygen overlap, and as limits the  motions in zeolites \cite{Thorpe2}}.
%
%
In order to prevent this in a computationally convenient manner, the harmonic potential is augmented with a purely repulsive potential
which acts between pairs of silicon atoms effectively acting as an inter-tetrahedron repulsive term.

\textcolor{black}{The physically more realistic imposition of a short-range repulsion between oxygen atoms requires greater computational accounting.}
The chosen form is a shifted 24-12 potential,

\begin{equation}
U(r)=4\epsilon\left\{\left(\frac{\sigma}{r}\right)^{24}-\left(\frac{\sigma}{r}\right)^{12}\right\}+\epsilon,
\label{equn:2412}
\end{equation}
where $\sigma$ is the atom diameter and $\epsilon$ is the well-depth of the (unshifted) potential. The potential
is cut off at the minimum [$r_{min}=(2)^{1/12}\sigma$] ensuring continuity in both energy and force.
The parameter $\epsilon$ is fixed
while $\sigma$ can be varied to explore the extent of the flexibility window.
In the second form a more realistic potential model (a TS potential \cite{tangney2002}) is used in which
pairwise-additive potential energy  terms (including ion-based charge-charge electrostatic interactions)
are augmented
with a description of (many-body) polarization
effects \cite{tangney2002,wilson1996,madden1996}.
This potential is more realistic than the harmonic potentials plus repulsions, mainly because Coulomb terms are included which are known to be important in ionic materials~\cite{Thorpe2} and we use this for a further optimization of the bilayer structure. Nevertheless the harmonic potential plus repulsions is useful as the language of flexibility windows and constraints and the symmetry plane can be used, as is discussed in the next section.

{\color{black}
We believe the choice of potential model is not crucial in displaying potentially interesting
phenomenology in systems of this type. The harmonic potential is chosen as (arguably) the simplest
model which constrains the system to form a series of ideal linked tetrahedral units. The TS potential is chosen as a potential
which accounts well for a number of key (bulk) properties whilst retaining a relatively
simple functional form.

Anion polarization,
which controls the Si-O-Si bond angles in models of this
type, may be crucial in
defining the structures adopted both for silica and potentially
for other, chemically-related, systems. Whilst the structures formed are low dimensional,
the atoms retain their full (bulk) coordination so it is reasonable (at least in the first approximation)
to apply potentials derived by reference to bulk three dimensional properties. These will most probably need further refinement as more precise experimental results on the vitreous silica bilayers become available.}

\textcolor{black}{We are very concerned with variation of the number density (number
of SiO$_2$ molecules per unit area) or, equivalently, the area occupied by a single molecule. The
number density, $n_0$, is expressed in terms of SiO$_2$ units per unit area projected onto the central plane of the bilayer, whilst the area, $A$, is expressed by reference to an ideal value, $A_0$,
which is the area occupied by a crystalline sample, based on crystalline graphene,
in which all the tetrahedral pairs are aligned vertically with a three fold axis about the central oxygen ion, and which would be obtained from bilayers constructed from an ideal crystalline graphene sheet containing only six-membered rings.}

Energy minimizations are performed over a range of dimensionless reduced
%
%
\textcolor{black}{areas $A^*=A/A_0;  0.4\le{A^*}\le{1.4}$.}
Note that $A^* =1$ is the maximum possible area
%
%
which can be attained without strain (i.e distortion within the individual tetrahedra).
At each density, the system's energy is minimised
using a steepest descent method. The atom positions are allowed to evolve, controlled by standard
Newtonian mechanics, and the velocities are reset to zero (quenching the kinetic energy) when the
kinetic energy reaches a (local) maximum.
\textcolor{black}{In order to allow explicitly breaking of the initial imposed reflection symmetry between the bilayers, several simulations are performed with randomized starting locations.}

\section{Results}
\begin{figure}[ht]
\includegraphics[width=70mm]{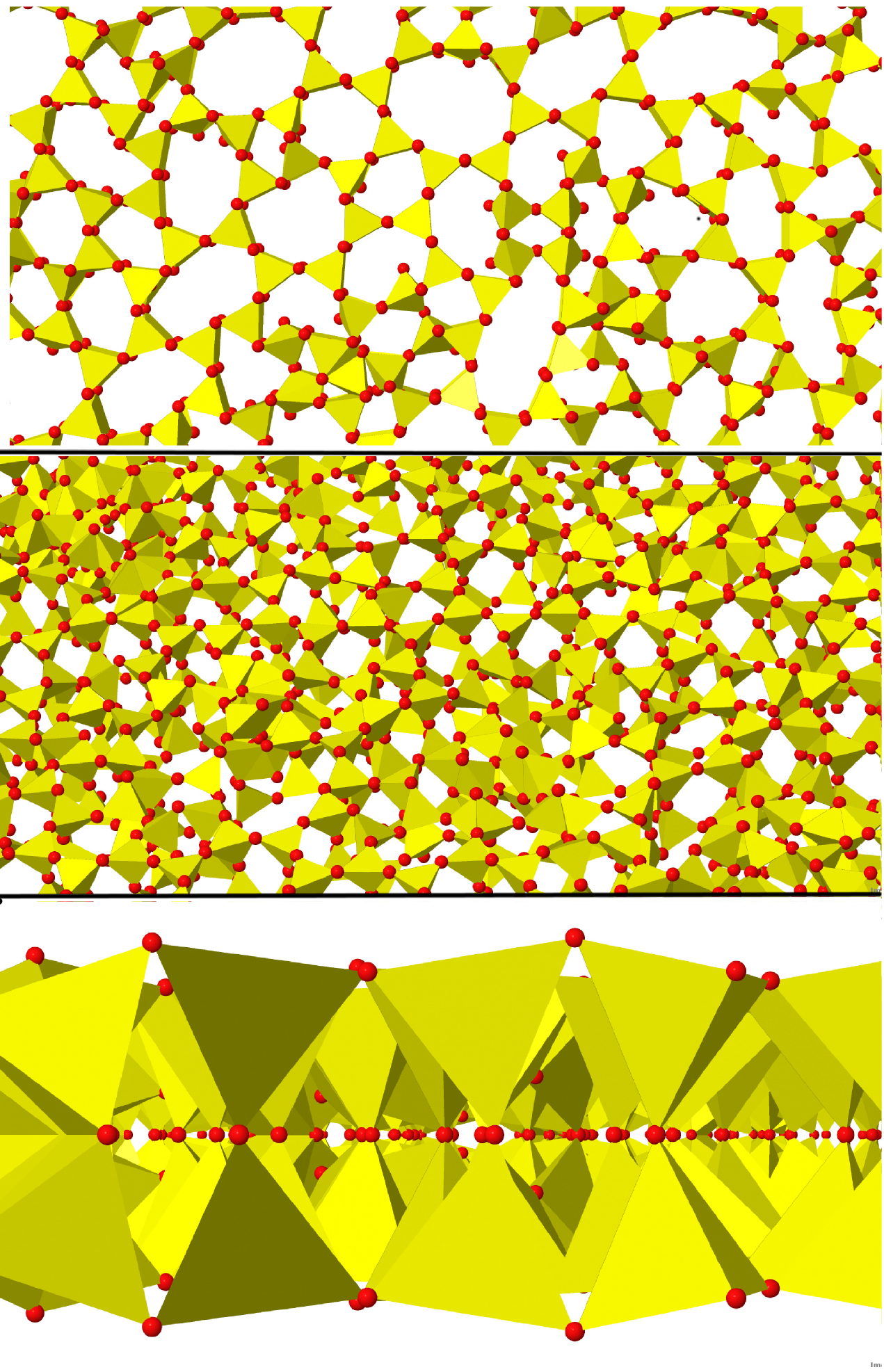}
\caption{Showing the structure of the bilayer made of perfect corner sharing tetrahedra from various perspectives, relaxed with
the harmonic potential and with a dimensionless area of $A^* = 0.9$. This puts the structure into an arbitrary place in the flexibility window at some density (not the experimental density). This illustration is a piece from a bilayer with a periodic super-cell with $864$ silicon atoms.
Notice that the central oxygen atoms all lie in the symmetry plane. Each tetrahedron has four oxygen ions shown as small red spheres at the vertices and a
silicon ion at the center.}
\label{fig:structure1}
\end{figure}
\begin{figure}[h]
\centering
\includegraphics[angle=0,scale=0.30]{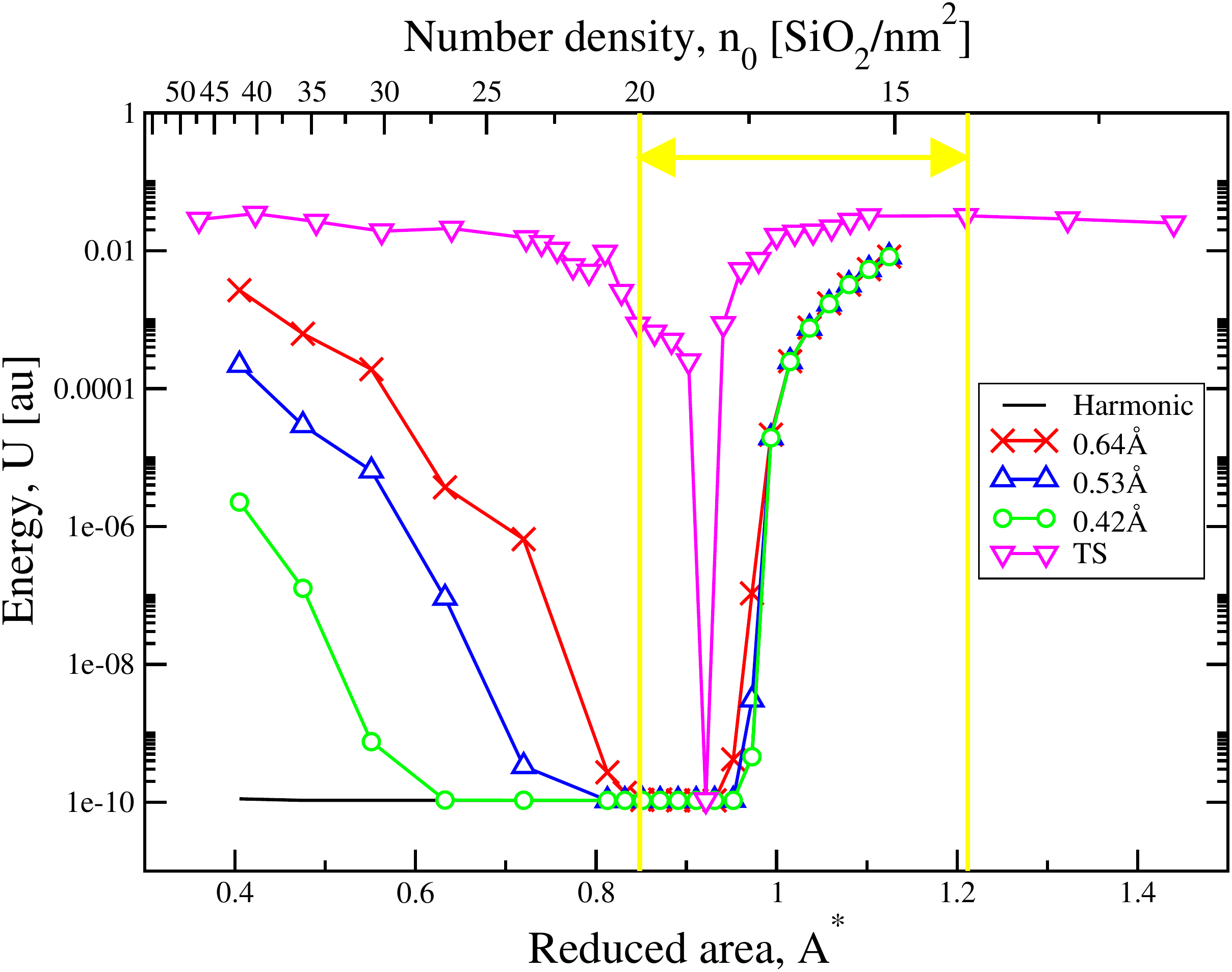}
\includegraphics[angle=0,scale=0.30]{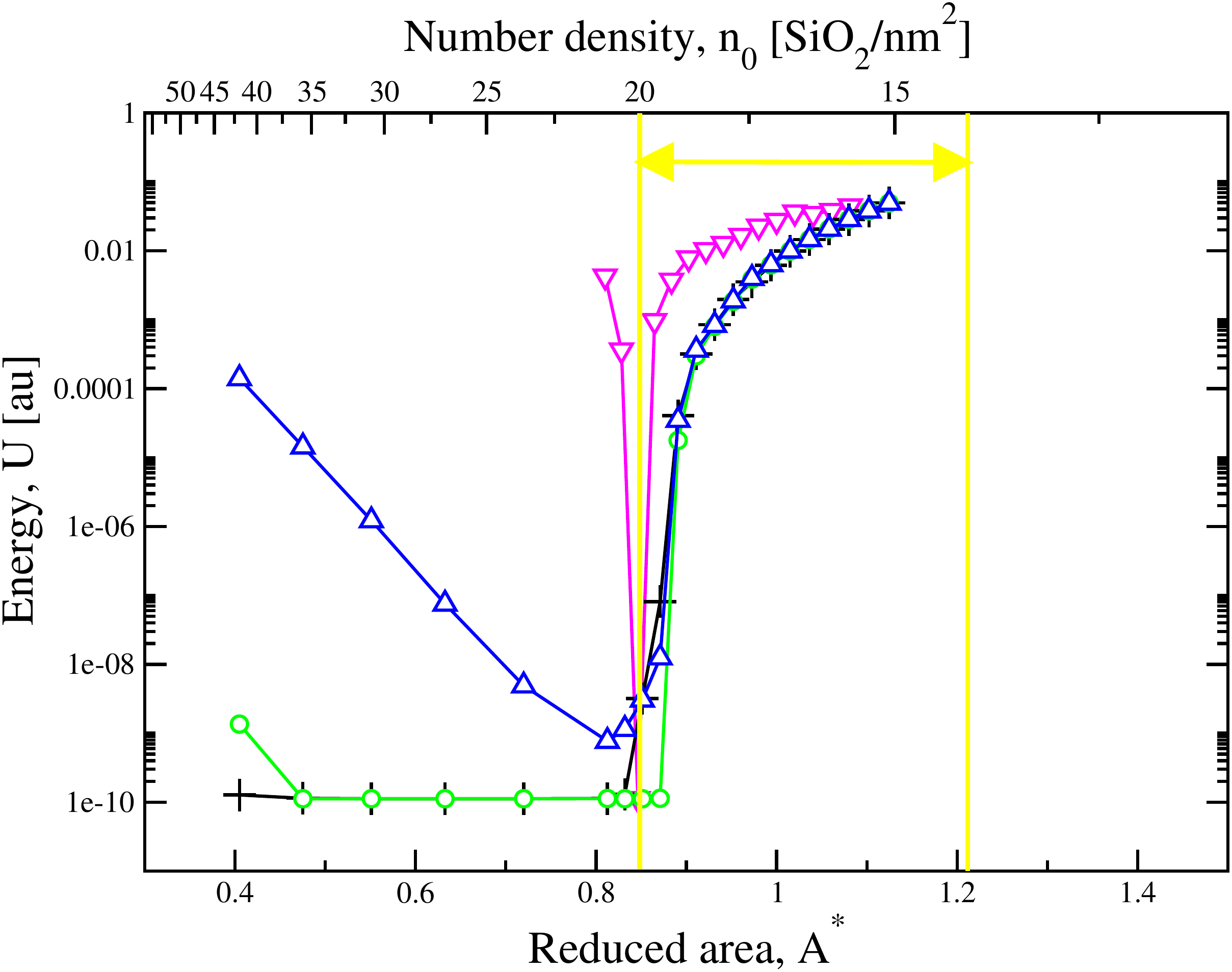}
\caption{The energies of the relaxed configurations shown as a function of the  density (upper abscissa)and the reduced
area, $A^*=A/A_0$ (lower abscissa) obtained for the 400 (upper panel) and 864 (lower panel)  SiO$_2$ molecule system,  using both harmonic
and TS potentials.
The distances indicated are for the parameter $\sigma$ in the box in the upper panel.
The minima for the TS potentials for the 400 and 864 molecule configurations are at densities of $\sim18.4$ SiO$_2$nm$^{-2}$ and
$\sim20.0$ SiO$_2$nm$^{-2}$ respectively.
The yellow lines and arrows in both panels
highlight the density range observed from experiment {\protect{\cite{cornell_pc}}}.}
\label{fig:U2}
\end{figure}

\textcolor{black}{In Figure~\ref{fig:U2}, we show comprehensive results for the harmonic and the, TS potential. These results are given for two different samples (derived from the 200 and 432a atom a-G samples, so containing a total of 1200 and 2592 atoms respectively), in the two panels to give some perspective on the universality of the results. These two configurations
are chosen as examples of systems with relatively high and low ring
distribution variances (Table \ref{tab:mu2}).
The results for the harmonic potential that describes the corner sharing network of rigid tetrahedra are shown by the red, blue and green lines and show a distinct flat region for both samples that is the manifestation of the flexibility window. These three curves are generated by different values of $\sigma$, with the smaller  values of $\sigma$, leading to larger flexibility windows. The flexibility window exists over a similar range of densities for both samples. It should be noted that the high density limit of the flexibility window is defined by repulsive potentials between the Si ions in this model, rather than the more physical repulsion between the larger O ions that is expected physically. However repulsion between the Si ions is expected to closely approximate the O-O repulsion, as the tetrahedra are all rigid.} The low density limit, defined as the lowest the density can be without breaking the network of corner sharing tetrahedra, is where almost all zeolites are found experimentally. This is because when a more realistic potential than the harmonic potential is used, Coulomb inflation maximises the pore volume, and hence the sample volume~\cite{Thorpe2}. The high density limit in zeolites is determined by that density at which interpenetration of the oxygen atomic spheres would onset.
A similar situation is found here for vitreous silica bilayers, with a well defined flexibility window.
This is in
%
%
contrast to the case of three dimensional vitreous silica (no pores) where the flexibility window collapses to a single point (single density)~\cite{Thorpe2}.

Figure~\ref{fig:U2} also shows the energies obtained by minimising the energy of the 400 and 864 SiO$_2$ molecular
bilayers using the TS potential.
 The energies resulting from the use of this
potential function show sharp minima (when plotted on a logarithmic scale used here).
These potentials produce a unique conformational minimum just below the high area limit of the flexibility window (obtained with the harmonic potential).
%
%

Figure~\ref{fig:U2} also shows the energies plotted against density; the harmonic potential results
are scaled by the Si-O bond length (1.6${\rm\AA}$). Also shown is the density
range obtained from experiment \cite{cornell_pc}.
For both configurations, studied with all potentials, the harmonic
potentials predict structures of slightly higher density than those observed preliminarily
experimentally.
On the other hand,
 for both these configurations the
energy minima for the TS potential do lie within the
currently observed
 experimental
density range ($n_0=18.4$ and $20.0$SiO$_2$/nm$^2$ for the 400 and 864 molecule
configurations respectively). The higher densities possible  using the harmonic
potential compared to the TS are
to be
expected
due to the lack of electrostatic
interactions which act to push the silicon cations apart. It is significant
to note that the two configurations studied in depth produce
energy minima with the TS at {\it different} densities implying
the density to be a function of the atomistic detail (ring structure)
of the bilayer configuration.

\begin{figure}[hbt]
\centering
\includegraphics[angle=0,scale=0.30]{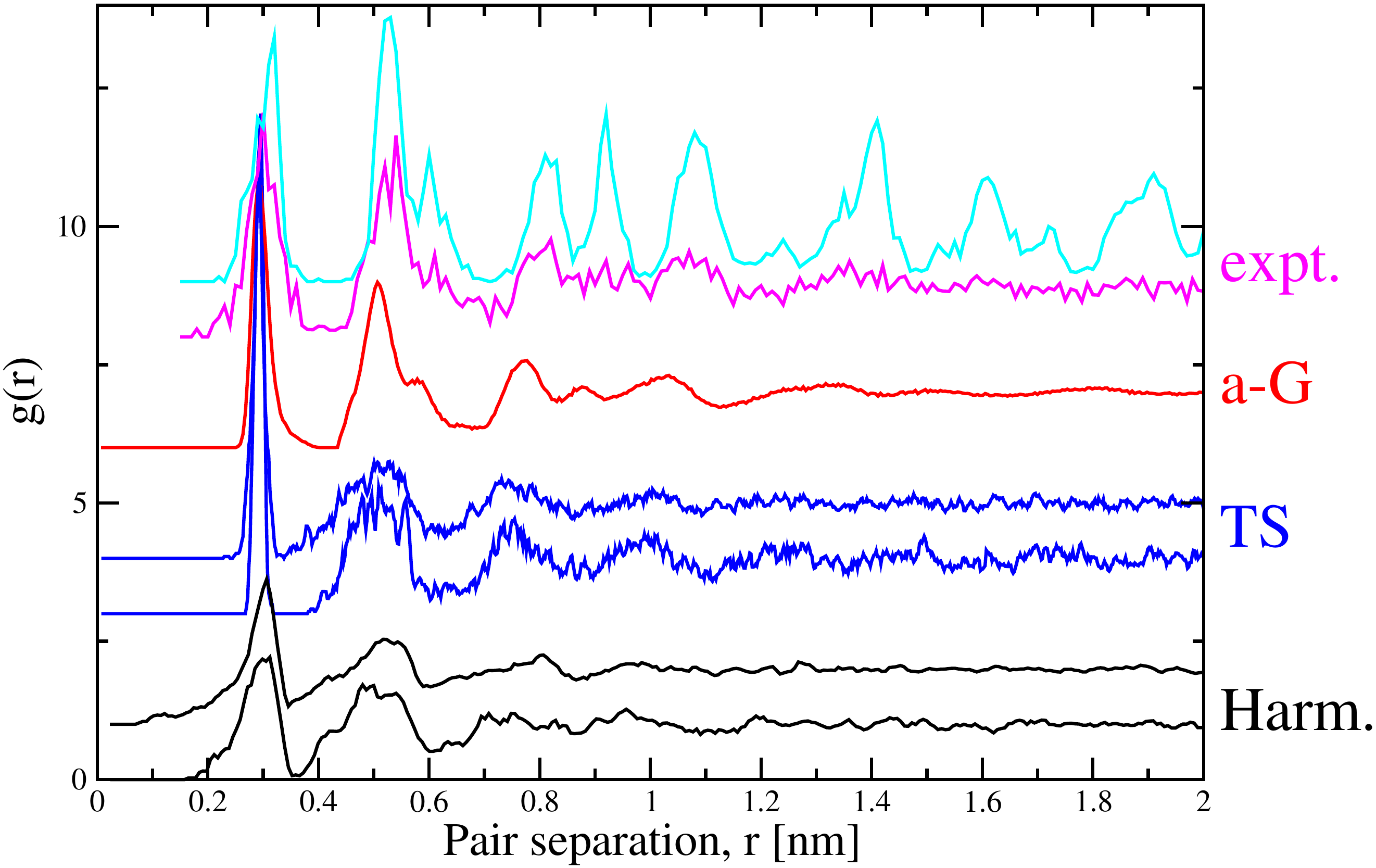}
\caption{Radial distribution functions $g(r)$ calculated for the silicon sublattice using the Si-Si
separations projected onto the $xy$ plane. The RDFs are calculated for the 400 and 864 SiO$_2$ molecule systems (respectively $\mu_{2}= 0.500$ and $ 1.046$)
using the harmonic (black lines) and TS potential (blue lines). Successive
functions are offset along the ordinate axes for clarity. In each case the lower curve is
for the 400 molecule configuration and the upper for the 864 molecule case.
The original carbon RDF determined from the a-G sample is also shown
(red).
 This function has been scaled
along the abscissa in terms of the first peak positions for comparison. The
harmonic potential functions are obtained at a density of $A^*=0.81$ (at which the energy
can be quenched) while the TS functions are obtained at the respective energy minima.
The uppermost curves are the amorphous (magenta) and crystalline (cyan) functions
obtained from experiment {\protect{\cite{cornell_pc}}}.}
\label{fig:rdf}
\end{figure}

Figure~\ref{fig:rdf} shows the Si-Si Radial Distribution Functions (RDFs) obtained
for both the harmonic and TS potentials for both the 400 and 864 molecule configurations. The RDFs are calculated by {\it{projecting}}
the Si-Si separations onto the $xy$ plane (and hence mimicking the experimental
procedure). 
 Energy minimisation using the TS potential produces structures with
order beyond the nearest-neighbor length-scale of the same form as that generated
by the harmonic potential. The first peak (corresponding to the nearest-neighbor
Si-Si length-scale) appears considerably sharper for the TS potential,
reflecting the ordering imposed by the presence of the electrostatic interactions.
The reduced intensities in the $864$ molecule system compared with the $400$ molecule
 one
reflects
the higher degree of disorder in the former (as
characterised
 by their respective values of $\mu_2$).

Figure~\ref{fig:rdf} also shows the experimentally-determined functions
(from Reference \cite{cornell_pc}) obtained for both a crystalline and an amorphous
section of bilayer.
The analysis of the experimental data remains very preliminary, and a much better determination
of the density will be possible once larger areas of the samples are imaged.
For the moment the fairly wide estimates of the experimental density~\cite{cornell_pc} are shown by the yellow lines in Figure~\ref{fig:U2}. These wide estimates are obtained from the relatively
small field of view of the vitreous silica bilayers currently available, and we await larger fields of view from which a more accurate density can be obtained. Note this density is obtained directly from the atomic imaging~\cite{cornell_pc}.
The density is a very important parameter to know, both in regards to the flexibility window and for detailed validation of the
potentials used here. It is quite possible that the potentials we have used will have to be fine tuned later to reflect the
experimental density
\textcolor{black}{but for the present we are concentrating principally on the conceptual physics.}
The preliminary experimental RDF shown in Figure~\ref{fig:rdf} is broadly consistent with all the model structures
in this paper, and data from much larger experimental areas should discriminate between the nuances of various computer generated structures.

\begin{figure}[h]
\includegraphics[angle=0,scale=0.30]{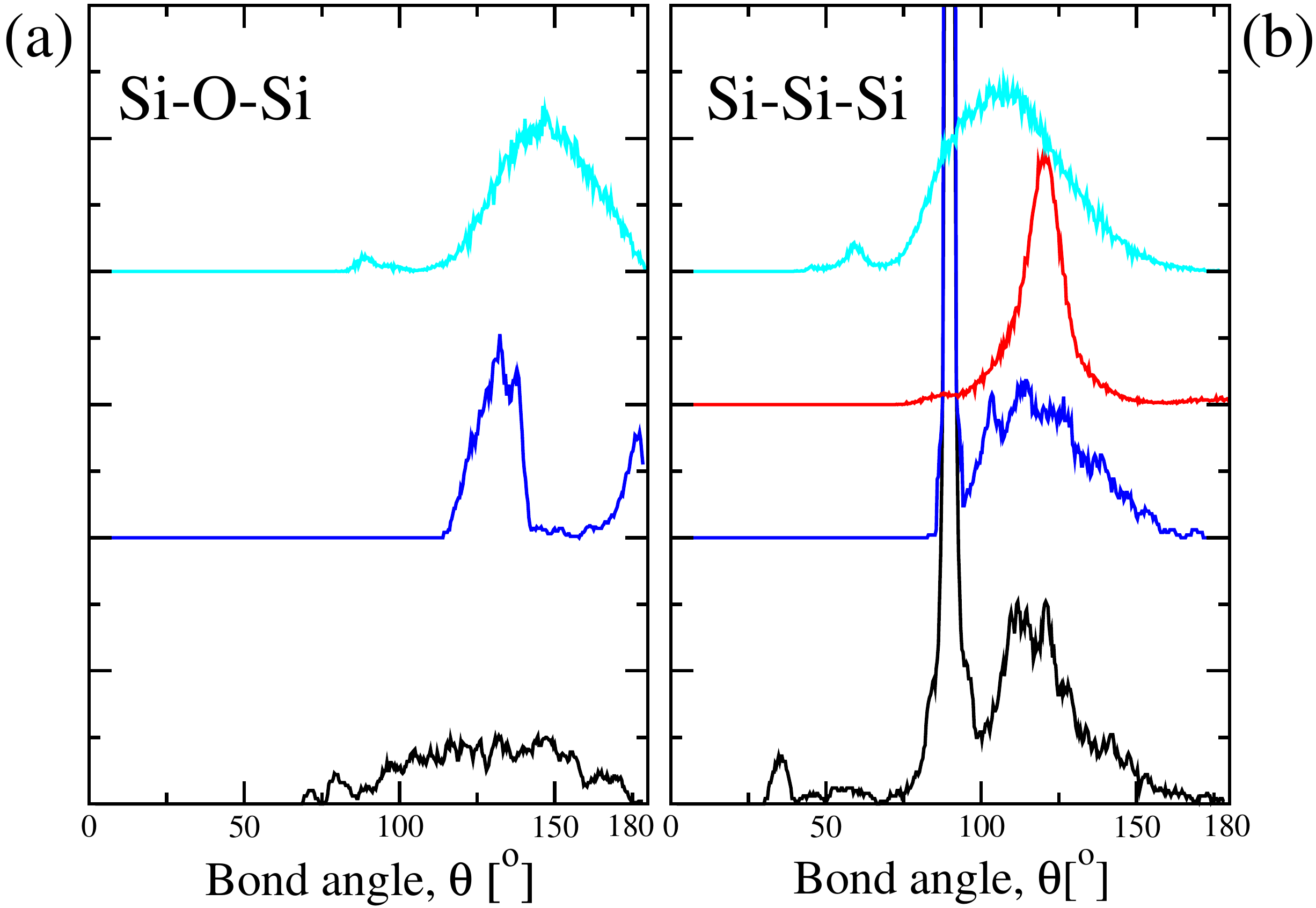}
\caption{(a) Si-O-Si bond angle distributions determined at a density of $A^*=0.90$
for the harmonic potential (black) and at the densities corresponding to the respective
energy minima for the TS (blue) for the 400 molecule system.
The light blue
plot
shows the corresponding function for the bulk glass at ambient pressure
using the TS potential
{\protect{\cite{wezka2013}}}.
(b) Si-Si-Si bond angle distributions
obtained under the same conditions as for panel (a). For reference the additional red line shows
the C-C-C distribution from the original two dimensional a-G configuration.}
\label{fig:bad}
\end{figure}

Figure \ref{fig:bad}(a) shows the Si-O-Si Bond Angle Distributions (BADs) determined for
the harmonic potential (at $A^*=0.81$ for which the energy could be driven to zero),
and at the energy minimum determined from the TS potential {\color{black}for the 400 molecule
system}.
The Si-O-Si BAD determined with {\color{black} either form of} the TS potential is significantly narrower than that determined from the
harmonic potential, reflecting the higher degree of ordering imposed
by the presence of the electrostatic interactions. The harmonic distribution shows a major peak at $\theta\sim{125}^o$
of width $\Delta\theta\sim$25$^o$ while the TS
shows a peak position (width)
of $\theta\sim$132$^o$ ($\Delta\theta\sim$15$^o$).
Note that the removal of anion polarization terms from the TS potential (to generate
a rigid-ion model) results in a
peak position (width)
of $\theta\sim$150$^o$ ($\Delta\theta\sim$10$^o$).
The change in the peak position in the bond angle distribution
is consistent with the inclusion of anion polarization which acts
to effectively screen the Si-Si (repulsive) electrostatic interactions and hence stabilises more
acute Si-O-Si bond angles.
The distribution appears
very different from that obtained for the bulk glass using the {\color{black}TS} potential \cite{wezka2013}
%
(also shown in Figure~\ref{fig:bad}) which shows a broader distribution ($\Delta\theta\sim$50$^o$)
with a peak at $\theta\sim{150}^o$ consistent with experiment \cite{neuefeind1996}.
\textcolor{black}{Note that in an unrelaxed crystalline bilayer, based on a crystalline graphene seed with a median symmetry plane, the Si-O-Si bond angle is cos$^{-1}(-7/9)${\color{red}$\simeq$}142$^o$ within a monolayer and 180$^o$ between monolayers. It is interesting to note that 142$^o$ is very close to the chemically preferred Si-O-Si bond angle in the absence of any topological {\it{strains}} due to rings~\cite{navrotsky}. The component at lower angles in the middle blue panel in Figure~\ref{fig:bad}(a) is associated with in-plane angles and the other peak with angles involving both planes. In the other two panels in Figure~\ref{fig:bad}(a), there is only a single very broad peak.}
%

\begin{table}[ht]
\begin{center}
\begin{tabular}{|c||c|c|c|c|c|}
\hline
$2N$ & Potential & $A^*$ & \multicolumn{3}{c|}{$\Delta{z}$/${\rm\AA}$} \\
\hline
&  & & upper & lower & center \\
\hline
400 & harm. & 0.81 & 0.124 & 0.124 & 3.9$\times$10$^{-14}$   \\
    & harm. & 1.00 & 0.087 & 0.087 & 2.8$\times$10$^{-14}$   \\
864 & harm. & 0.81 & 0.184 & 0.184 & 4.0$\times$10$^{-14}$   \\
    & harm. & 1.00 & 0.137 & 0.137 & 3.9$\times$10$^{-14}$   \\
\hline
400 & TS & 0.92 & 0.151 & 0.151 & 7.2$\times$10$^{-14}$   \\
864 & TS & 0.85 & 0.274 & 0.274 & 9.0$\times$10$^{-14}$   \\
\hline
\end{tabular}
\end{center}
\caption{Variances in the position of the oxygen atoms perpendicular
to the bilayer plane, $\Delta{z}$, for the 400 and 864 SiO$_2$
molecule configurations determined using both the harmonic and TS potentials
at the reduced areas indicated. The oxygen atoms sit in three layers;
 upper, lower and central.
 The
small variances for the atoms in the central plane (of the order of the numerical error associated
with the calculation), coupled
with the identical variances of the atoms above and below
this plane are indicative of the existence of a mirror plane
containing the central oxygen atoms.}
\label{tab:mirror}
\end{table}

The difference between the bilayer and bulk BADs can be rationalised as follows. The mirror symmetry
relationship between the top and bottom layers of the bilayer means that ions of the same charge sit
on top of one another perpendicular to the plane containing the bilayer. As a result, the (repulsive)
like-like electrostatic interactions are effectively maximised, leading to relatively obtuse
Si-O-Si bond angles centred about the oxygen ions which bridge the two layers, and resulting
in the peak at $\sim$175$^o$. A simple geometric argument indicates that the presence of these relatively
obtuse angles has a knock-on effect for the Si-O-Si angles centred about the oxygen ions which are in
one of the bilayer planes, which will be relatively acute. This is an area that needs more study as %
there will always be
 a
competition between the preferred Si-O-Si angles from chemistry and the requirements of the network topology. This effect will influence
whether similar vitreous bilayers can be made from germanium and also whether
aluminum
 ions can be alloyed with
silicon
ions in these
vitreous silica bilayers.

Figure \ref{fig:bad}(b) shows the Si-Si-Si bond angle distributions obtained using
the harmonic and the TS potentials
at the same densities as in Figure \ref{fig:bad}(a). The results are compared to the C-C-C BAD
generated from the original a-G configuration.
Recall that for crystalline graphene the bond angle is $\theta\sim$120$^o$.
 The bilayer BADs show a sharp
peak at $\theta\sim$90$^o$ which corresponds to Si-Si-Si triplets in which the
Si atoms are split between the two layers comprising the bilayer,
while the higher broader peak corresponds to in-plane Si.
The harmonic and TS
potentials show similar distributions which are significantly broader than
the a-G distribution.


To quantify the presence of a mirror plane
after relaxation, which does not impose any symmetry,
we determine the variance of
the oxygen atom positions ($\Delta{z}$) perpendicular to the bilayer plane. The
oxygen atoms can be considered as sitting in three distinctive
quasi-planes
corresponding to the central layer (which {\it{joins}} the two
original monolayers) and the two layers above and below this central
layer. Table \ref{tab:mirror} lists the variances for the
400 and 864 SiO$_2$ molecule systems obtained using the
harmonic potential (at two densities) and the TS
(in the respective energy minima densities). The central
atoms are clearly confined to a single plane while the atoms
in the upper and lower layers show identical variances. The
existence of the mirror plane is confirmed by determining
the variance in the positions of the mirrored atoms in the upper
and lower planes which is $\sim{0}$ to within the numerical precision
used. In principle the mirror symmetry
%
%
can
be broken once the potential contains Coulomb terms etc.,
as the argument given in the section {\em{Symmetry Planes}} based upon constraints does not hold with more complex forces.
%
%
However in practice it seems these deviations are very small, although in principle present. We also note that this symmetry in the $\Delta{z}$ perpendicular to the bilayer plane, holds not only at the macroscopic 9(average) level, but also at the local level between corresponding atoms above and below the central symmetry plane.

\begin{figure}[h]
\centering
\includegraphics[width=70mm]{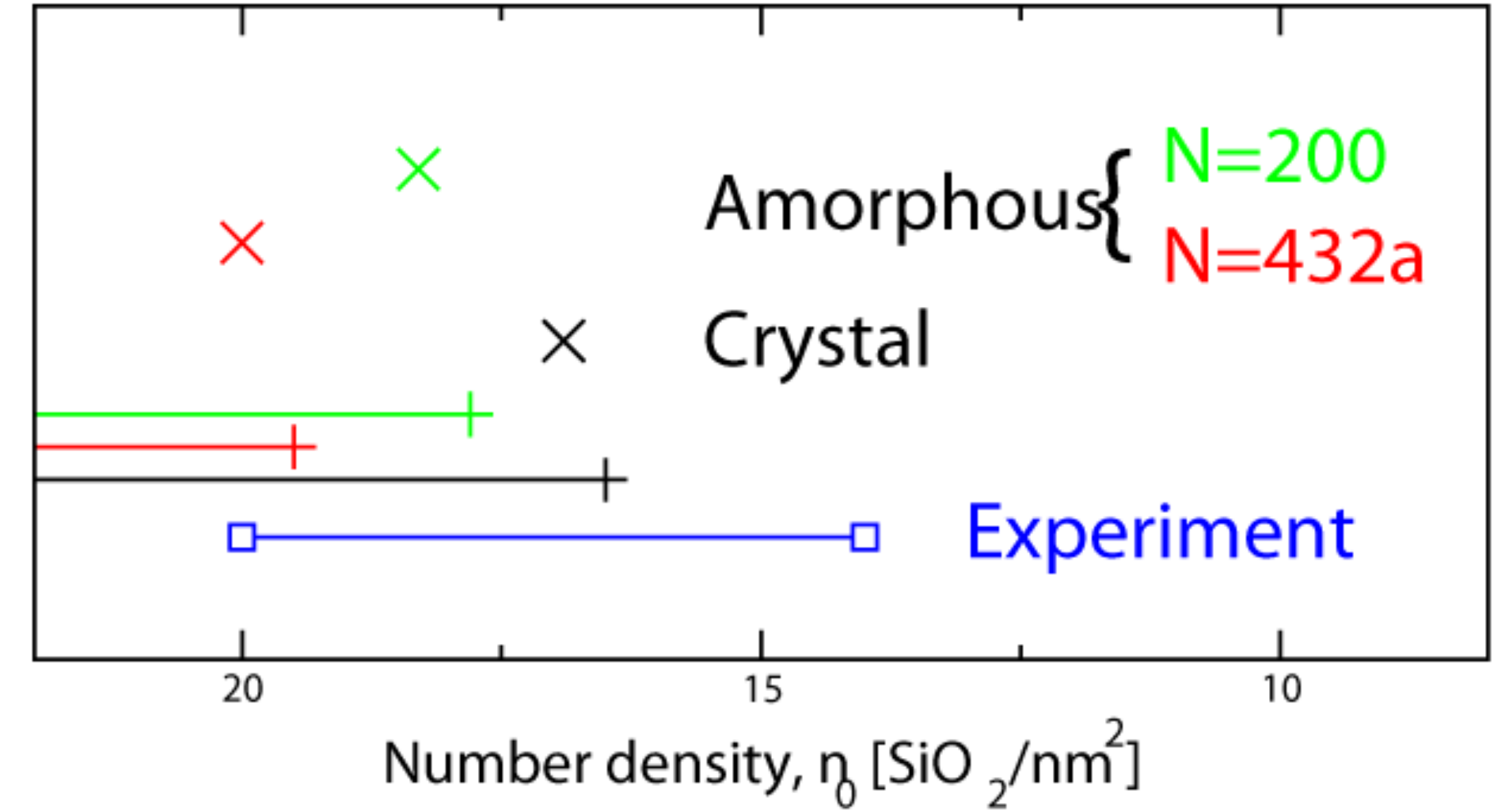}
\caption{The low density end of the flexibility window (for the harmonic potential case) for the two amorphous (green and red - sample
sizes as indicated) and one crystalline
(black) sample shown as lines ending in a $+$ symbol. Note that the The high density end of the window for harmonic potentials is to the left and determined by repulsive forces between neighboring ions.  The figure also shows the corresponding energy
minima from the TS potential ($\times$) coloured as for the corresponding flexibility windows.
The blue line highlights the experimental density range.}
\label{fig:summary}
\end{figure}

The most important results of this section, and paper, are summarised in Figure ~\ref{fig:summary}
where the flexibility windows and corresponding energy minima for the two amorphous and one crystalline sample are shown.
It should be noted that the TS minima lies within the flexibility window as expected.
In addition it lies towards the top, high area end, of the window reminiscent of the relationship
observed in $3d$ zeolites~\cite{Thorpe5}. The argument given for the zeolites was that Coulomb inflation
in the pores between the negative oxide ions caused the sample to swell to be very close to the maximum allowed
while remaining inside the flexibility window. Such an argument cannot be given here, as there are no large pores
as in zeolites - but we propose that Coulomb inflation, between the oxide ions, may still be the explanation within the rings of the
bilayer. The silicon ions are less important as they are smaller.
However we do not find this argument entirely convincing, and more work on understanding the subtleties of the
effects of Coulomb interactions in ionic framework structures is needed.

\section{Symmetry plane}

The existence of a symmetry plane in an amorphous sample is surprising and quite unlike anything that we have encountered before.  This symmetry emerges from the disordered state as the network takes advantage of the larger conformational space available when a symmetry plane is present compared to without. Hence symmetry is induced in a system which at first sight seems a canonical example of a system without symmetry. The argument for it is compelling as outlined here, and confirmed both by detailed atomic computer modeling and by experiment.  Of course the individual tetrahedral units are close to perfect tetrahedra because of the strong local chemical bonding, but these pack in a disordered way. The local symmetry of perfect  tetrahedra is not necessary for the argument for the symmetry plane to hold. Residual degrees of freedom in a structural unit are often referred to as floppy modes~\cite{Thorpe}.

\begin{figure}[ht]
\centering
\includegraphics[width=80mm]{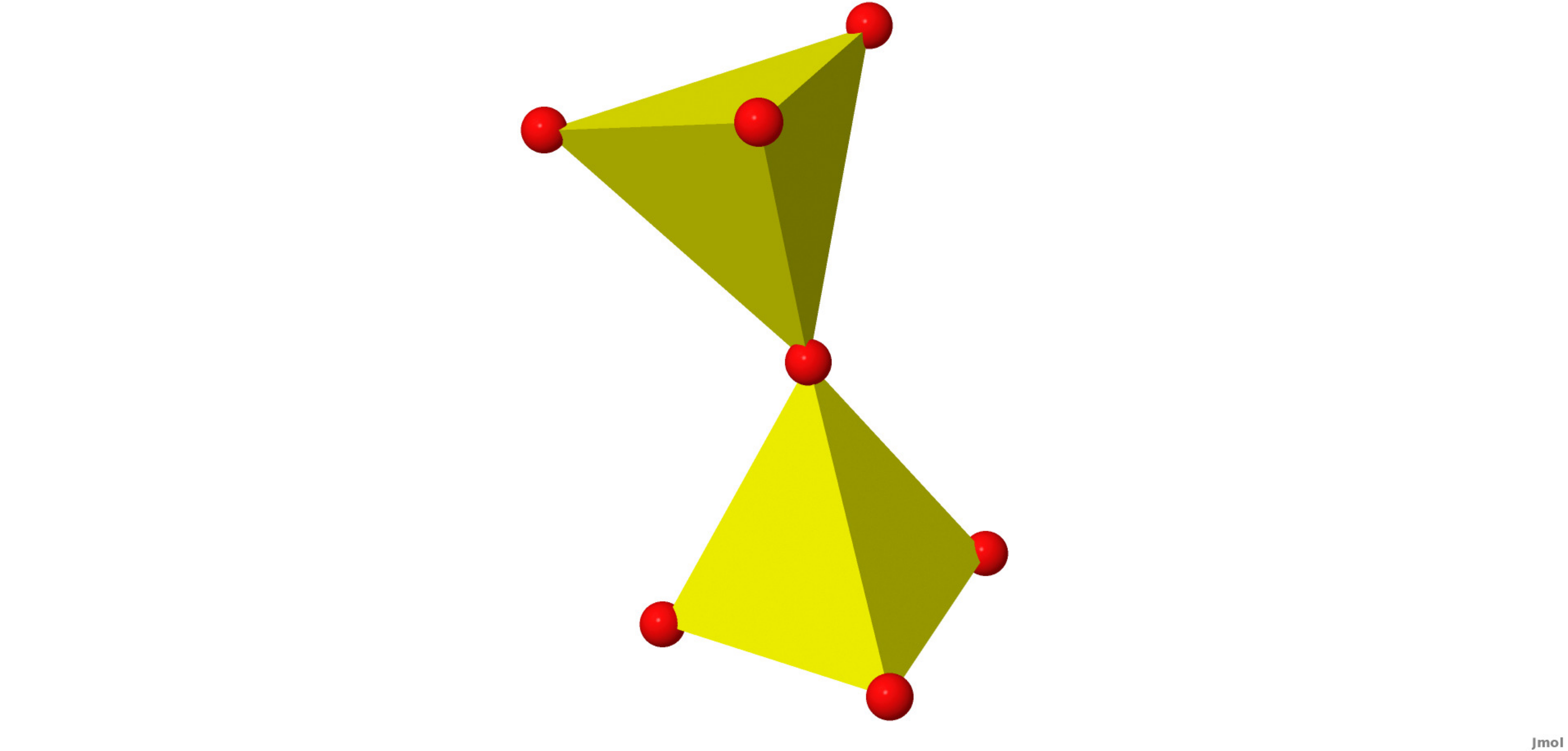}
\caption{Showing a tetrahedral pair; one from the upper layer and one from the lower layer. Each tetrahedron has four oxygen ions shown as small red spheres at the vertices and a silicon ion (not shown) at the center.}
\label{fig:structure}
\end{figure}

Consider two tetrahedral units (which we will call a {\it tetrahedral pair}) with a common oxygen atom  - that is Si$_{2}$O$_{7}$ as shown in Figure~\ref{fig:structure}.  There are $N$ such units made from the $2N$ units of SiO$_{2}$ in the bilayer structure. A rigid body has 6 degrees of freedom in three dimensions, and for any two vertices of adjacent tetrahedra to coincide requires 3 constraints ( x$_{1}$ = x$_{2}$, y$_{1}$ = y$_{2}$ and z$_{1}$ = z$_{2}$, where 1 and 2 refer to the two tetrahedral vertices  that come together). We now give counting arguments~\cite{Thorpe} for the case without symmetry and then with symmetry between the bilayers from a (proposed) reflection plane in the middle of the two layers.

1.	{\em Without symmetry}.  Consider the two separate tetrahedra in the tetrahedral pair shown in Figure~\ref{fig:structure}; each with the usual 6 degrees of freedom for a rigid body in three dimensions. That makes 12 degrees of freedom in total.  Joining the common oxygen atom requires 3 constraints (to make the tetrahedral pair) and the 6 remaining oxygen ions each require 3 constraints, which are shared so that there are ${12-3 -(6\times 3)/2 = 0}$ floppy modes, which is the expected result that the system is isostatic~\cite{Thorpe2} - that is the structure just has enough constraints to be rigid and cannot be moved (subject to the usual remarks about boundary conditions {\it etc}.~\cite{Thorpe4}). This result is expected as the bilayer is just a special case of a network of corner sharing tetrahedral units which are all isostatic~\cite{Thorpe2}. This can be seen easily as each individual isolated tetrahedron has 6 degrees of freedom, and four shared corners each with 3 constraints - hence the number of floppy modes per tetrahedron is ${6 - (4 \times 3)/2 = 0}$ and we have an isostatic network; that is no floppy modes.

2.	{\em With symmetry}.  First define an external plane and position it at $z=0$. Then move a single tetrahedron so one vertex lies in this plane but is free to move in the $x-y$ plane.  This tetrahedron initially has 6 degrees of freedom, and putting an atom somewhere in the external plane requires a single constraint. The 3 other oxygen  atoms require $(3 \times 3)/2$ constraints to link the corners - the second tetrahedron in the tetrahedral pair is fixed by the reflection symmetry and therefore has neither independent degrees of freedom nor independent constraints associated with it. Thus there are ${6-1-(3\times 3)/2 = 1/2}$ remaining degrees of freedom for each tetrahedral pair and hence a total of $N/2$ degrees of freedom for the whole bilayer.

The important observation here is that with corner sharing tetrahedra, there are a macroscopic number of degrees of freedom {\em if} there is a reflection plane.  This allows the bilayer structure {\it to roam over a region of conformational space} while maintaining all the constraints.  This leads to a flexibility window analogous to that found previously in zeolites~\cite{Thorpe2}.  As discussed later, we find that with a more realistic potential the bilayer settles within the flexibility window at a preferred density.   This is a pretty remarkable and unexpected result.

The above arguments can be streamlined and made very compact. Looking at the lower monolayer, disconnected from the upper monolayer, then
if
there is a reflection plane only a single constraint is required at \textcolor{black}{each}
oxygen that is not shared with another tetrahedron
\textcolor{black}{in order
to bring it to the reflection plane, whereas three shared constraints
%
%
would be required to bring it into coincidence with a similar oxygen in an otherwise unconstrained second monolayer.
Adding the mirror monolayer adds no extra degrees of freedom.}
Therefore
a reflection plane leads to
a macroscopic number of floppy modes $(3/2 -1)N = N/2$.

Yet another distinct demonstration of this result can be given as follows.
Start with planar collection of $N$ corner-sharing triangles. There are $3N$ degrees of freedom and $3\times 2N/2$ constraints giving no remaining degrees of freedom.  That is such a network (example is a kagome lattice) is isostatic and only flexible if there is a surface  and (O${\sqrt{N}}$) floppy modes.
 Now consider a $3d$ flexible framework of tetrahedra connected with the same $2d$ topology of corner sharing triangles (each tetrahedron has three connected corners), then there are $6N$ degrees of freedom and $3\times 3N/2$ constraints giving a macroscopic number of degrees of freedom $3N/2$.   If further, the remaining vertex of each tetrahedron, so fixing one of its Cartesian coordinates, then the total number of degrees of freedom per tetrahedron is $N/2$.  Adding the reflected monolayer to the original monolayer gives $2N$ SiO$_{2}$ units with $N/2$ degrees of freedom for the bilayer, as before.

Note that these arguments apply to tetrahedra of arbitrary shape and size, so if there are some Al ions within tetrahedra in the lower layer at some composition, they should be mirrored in the upper layer based on the arguments in this sections, although Coulomb repulsions between them  would discourage this. Aluminum and silicon tetrahedra are both nearly perfect but have different sizes~\cite{Thorpe2} and so attempting to construct such bilayers
would be something
%
%
interesting to try experimentally. The situation is very different in
conventional
zeolites, where Loewenstein's
empirical rule~\cite{Lowen} for alumino-silicates
%
states that (in essence) every
\textcolor{black}{Al - containing tetrahedra
must be connected to four neighboring
silicon-containing
 tetrahedra.
 It is possible, therefore, that Loewenstein's rule
%
%
%
could
%
%
sometimes be violated in bilayer structures.}

\section{Conclusions} We have shown how the recently discovered vitreous silica bilayer can be computer-modeled by progressive assembly, starting from
an amorphous
 graphene sheet,
%
%
 and making various decorations
and then relaxation with appropriate potentials.
%
 This pathway is of course not physical, but represents a convenient way of computer-generating such structures.

This
system
 is probably the first network glass where the ring structure can be experimentally observed directly by STM and STEM measurements with atomic resolution, making
it
 a paradigm system for future study.
\textcolor{black}{This present study provides a complementary computer-theoretical study  that we hope will encourage further experimental and theoretical work.}

An interesting observation is the unexpected mirror symmetry plane through the center of the bilayer, which seems to defy the logic which says that such bilayers should pucker.  The fact that the top monolayer lies exactly on top of the lower monolayer, means that a single layer is seen in the experimental STEM image, making structural interpretations much easier, and confirming the symmetry argument given here. This is a very unusual situation of a symmetry induced in a disordered system.
\textcolor{black}{However it should be noted that such a reflection symmetry is also expected in crystalline silica bilayers, where it is less surprising.}
%
%
Although additional terms in the potential, like the Coulomb terms,
may result in this symmetry being broken, this is not observed in the experiments, or in the computer simulations, at the current level of accuracy. It should be noted that amorphous graphene shows no such symmetry and is expected to show considerable puckering, unless constrained from doing so by a sufficient applied tension~\cite{kumar2012}.

\section{Acknowledgments} We should like to thank Mike Treacy for many useful discussions including those relating to Lowenstein's rule and the US National Science
Foundation for support under Grant No.~DMR-0703973. We should also like to thank Pinshane Huang and David Muller for sharing their experimental data.
DS acknowledges support from the Leverhulme Trust through an Emeritus Research Fellowship.

\bibliographystyle{unsart}

\end{document}